\documentclass[letter]{aa} 

\usepackage{graphicx}
\usepackage{txfonts}
\usepackage[flushleft]{threeparttable} 
\usepackage{soul}
\usepackage{amssymb}
\usepackage{lscape}
\usepackage{comment}
\usepackage{ulem}

\usepackage{hyperref}
\hypersetup{colorlinks=true,linkcolor=blue,citecolor=blue,filecolor=blue,urlcolor=blue}

\usepackage{multirow}

\begin{document} 
\newcommand{\he}[1] {He\,{\sc #1}}
\newcommand{\hel}[2] {He\,{\sc #1}~{$\lambda$#2}}
\newcommand{\ha}{H$\alpha$}
\newcommand{\hb}{H$\beta$}
\newcommand{\hg}{H$\gamma$}
\newcommand{\hd}{H$\delta$}
\newcommand{\hep}{H$\epsilon$}
\newcommand{\hei}{He~{\sc i}}
\newcommand{\heii}{He~{\sc ii}}

%
%
\def\kms{\mbox{${\rm km}\:{\rm s}^{-1}\:$}}
\def\ms{${\rm m}\:{\rm s}^{-1}\:$}

\def\lesssim{\mathrel{\hbox{\rlap{\hbox{\lower4pt\hbox{$\sim$}}}\hbox{$<$}}}}
\let\la=\lesssim
\def\gtrsim{\mathrel{\hbox{\rlap{\hbox{\lower4pt\hbox{$\sim$}}}\hbox{$>$}}}}
\let\ga=\gtrsim
\def\fdg{\hbox{$.\!\!^\circ$}}
\newcommand{\fdeg}{\hbox{$.\!\!^\circ$}}
\def\farcs{\hbox{$.\!\!^{\prime\prime}$}}
\def\arcsec{\hbox{$^{\prime\prime}$}}
\def\kmm{${\rm km}^{-1}$}
\def\d{{\rm d}}
\def\degree{\mbox{$^{\circ}$}}
\def\sol{$~\mathrm{M}_\odot$}
\def\lx{$L_\mathrm{X}$}
\def\kem{K_\mathrm{em}}
\def\ergs{erg s$^{-1}$}
\def\ledd{$L_\mathrm{Edd}$}
\def\vt{$v_\mathrm{t}$}

\def\sloanu{$u^\prime$}
\def\sloang{$g^\prime$}
\def\sloanr{$r^\prime$}
\def\sloani{$i^\prime$}
\def\sloanz{$z^\prime$}

    \title{Fast infrared winds during the radio-loud and X-ray obscured stages of the black hole transient GRS~1915+105}

   \author{J.~Sánchez-Sierras\inst{\ref{i1},\ref{i2}}
          \and
          T.~Muñoz-Darias\inst{\ref{i1},\ref{i2}}
          \and
          S.~E.~Motta\inst{\ref{i3},\ref{i4}}
          \and
          R.~P.~Fender\inst{\ref{i4},\ref{i5}}
          \and
          A.~Bahramian\inst{\ref{i6}}
          \and
          C. Martínez-Sebastián\inst{\ref{i1},\ref{i2}}
          \and
          J.~A.~Fernández-Ontiveros\inst{\ref{i7}}
          \and
          J.~Casares\inst{\ref{i1},\ref{i2}}
          \and
          M.~Armas~Padilla\inst{\ref{i1},\ref{i2}}
          \and
          D.~A.~Green\inst{\ref{i8}}
          \and
          D.~Mata~Sánchez\inst{\ref{i1},\ref{i2}}
          \and
          J.~Strader\inst{\ref{i6}}
          \and
          M.~A.~P.~Torres\inst{\ref{i1},\ref{i2}}
          }

   \institute{
            Instituto de Astrofísica de Canarias, c/Vía Láctea, S/N, E-38205 La Laguna, Tenerife, Spain \label{i1}
        \and
            Departamento de Astrofísica, Universidad de La Laguna, E-38206 La Laguna, Tenerife, Spain \label{i2}
        \and
            Osservatorio Astronomico di Brera, Via E. Bianchi 46, I-23807 Merate (LC), Italy \label{i3}
        \and
            Department of Physics, University of Oxford, Denys Wilkinson Building, Keble Road, Oxford OX1 3RH, UK \label{i4}
        \and
            Department of Astronomy, University of Cape Town, Private Bag X3, Rondebosch 7701, South Africa \label{i5}
        \and
            International Centre for Radio Astronomy Research, Curtin University, GPO Box U1987, Perth, WA 6845, Australia \label{i6}
        \and
            Centro de Estudios de Física del Cosmos de Aragón (CEFCA), Plaza San Juan 1, 44001 Teruel, Spain \label{i7}
        \and
            Astrophysics Group, Cavendish Laboratory, 19 J. J. Thomson Avenue, Cambridge CB3 0HE, UK \label{i8}
             }

\titlerunning{Near-infrared winds in GRS~1915+105}
\authorrunning{Sánchez-Sierras et al. 2023}
 
  \abstract
  {
    The black hole transient GRS~1915+105 entered a new phase of activity in 2018, generally characterised by low X-ray and radio fluxes. This phase has only been interrupted by episodes of strong and variable radio emission, where high levels of X-ray absorption local to the source were measured. We present 18 epochs of near-infrared spectroscopy (2018--2023) obtained with GTC/EMIR and VLT/X-shooter, spanning both radio-loud and radio-quiet periods. We demonstrate that radio-loud phases are characterised by strong P-Cygni line profiles, indicative of accretion disc winds with velocities of up to $\mathrm{\sim 3000~km~s^{-1}}$. This velocity is consistent with those measured in other black hole transients. It is also comparable to the velocity of the X-ray winds detected during the peak outburst phases in GRS~1915+105, reinforcing the idea that massive, multi-phase outflows are characteristic features of the largest and most powerful black hole accretion discs. Conversely, the evolution of the Br$\gamma$ line profile during the radio-quiet phases follows the expected trend for accretion disc lines in a system that is gradually decreasing its intrinsic luminosity, exhibiting weaker intensities and more pronounced double-peaks.
    
  }


   \maketitle


\section{Introduction}

The black hole (BH) transient GRS~1915+105 has the longest orbital period among its class and, consequently, a large and powerful accretion disc ($P=33.85\pm0.16$ days; \citealt{Steeghs2013, Corral-Santana2016}). This low-mass X-ray binary (XRB) was discovered in 1992 \citep{Castro-Tirado1994} during the onset of an outburst. Since then, it has been active for more than 25 years, exhibiting a diverse range of accretion-related phenomena and associated outflows, such as jets and hot X-ray winds (e.g. \citealt{Mirabel1994,Eikenberry1998,Fender2004b, Neilsen2009, Neilsen2011}).

\begin{figure*}[ht]
    \centering
    \includegraphics[width=18truecm]{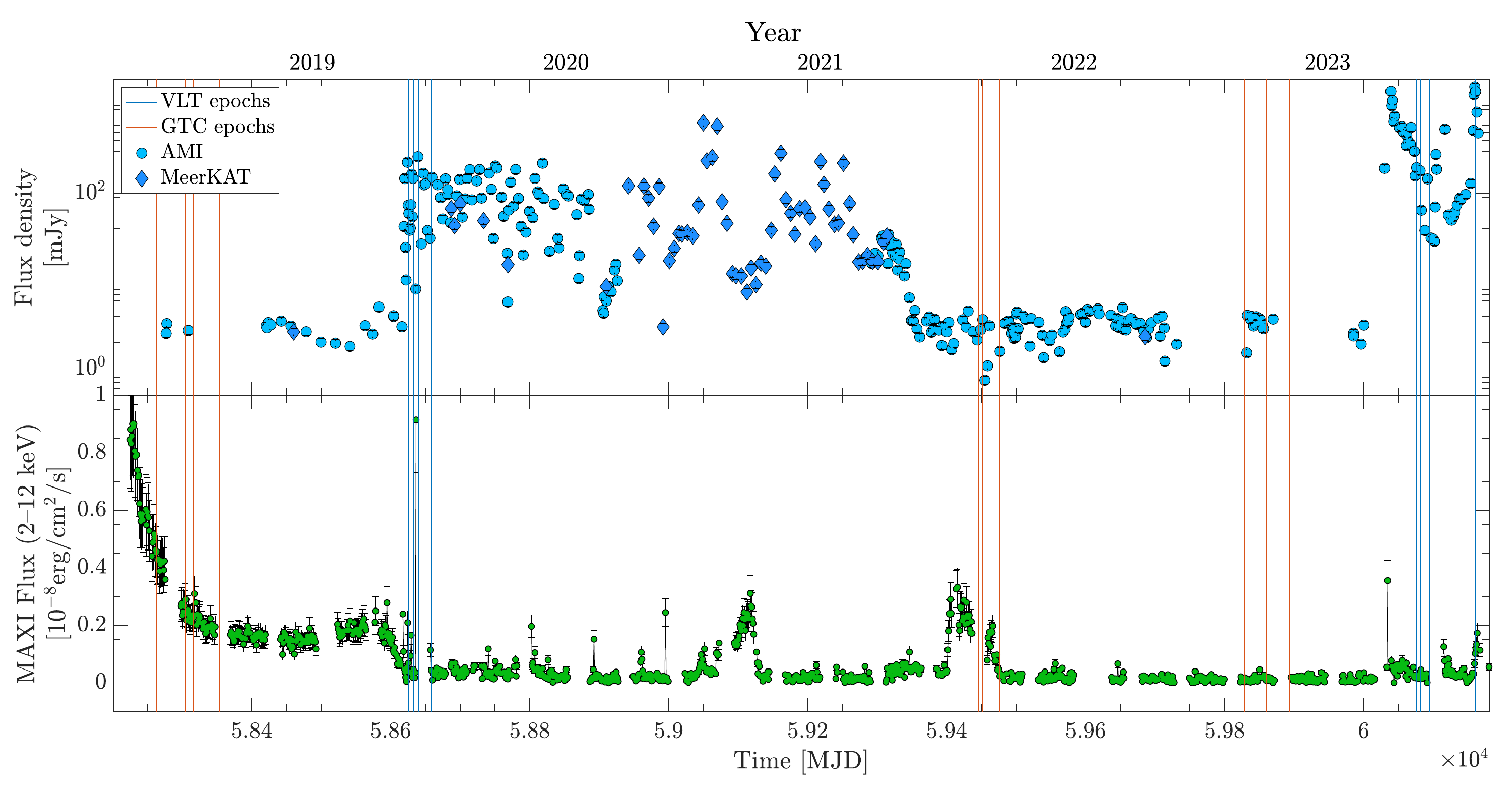}
    \caption{Light curves in radio (MeerKAT and AMI; top panel) and X-rays (MAXI; bottom panel) from March 2018 (MJD 58200) to August 2023 (MJD 60170). Vertical lines indicate the spectroscopic epochs of GTC/EMIR (orange) and VLT/X-shooter (blue).}
    \label{figLightCurves}
\end{figure*}

In 2018, GRS~1915+105 entered a new phase (see Fig. \ref{figLightCurves}). The X-ray and radio fluxes started to decrease gradually, reaching a plateau stage at low flux. This was followed (May 2019) by a further decline in the observed X-ray emission, which was surprisingly accompanied by an increase in the radio flux back to pre-2018 levels. Strong intrinsic obscuration was inferred from X-ray spectroscopy obtained during this phase, which could qualitatively explain (at least in part) the puzzling behaviour of the source \citep{Miller2020,Motta2021,Balakrishnan2021}. This radio-loud phase at low X-ray flux lasted more than two years and it does not have clear precedent in XRBs. Once this phase ended, the source displayed both low X-ray and radio fluxes, consistent with a gradual return towards quiescence. Finally, after several months of particularly low activity, the source started a new radio-loud phase in April 2023 \citep{Egron2023}.

GRS~1915+105 cannot be observed at optical wavelengths due to a very high extinction along the line of sight \mbox{($A_{V} = 19.6 \pm 1.7$}; \citealt{Chapuis2004}). Thus, the near-infrared (NIR) offers a unique observing window to study the outer accretion disc and the possible presence of low-ionisation (also known as cold) winds (e.g. \citealt{Munoz-Darias2019}; see also \citealt{Ponti2012, DiazTrigo2016, Parra2023} for hot, X-ray winds). These outflows are particularly conspicuous in other transient XRBs with large accretion discs (e.g. V404~Cyg and V4641~Sgr; \citealt{Munoz-Darias2016,Munoz-Darias2017, Munoz-Darias2018}) and can be associated with high levels of variable, intrinsic X-ray absorption (e.g. \citealt{Motta2017a, Motta2017b}). Here, we present near-infrared spectroscopy and photometry of GRS~1915+105 covering the different phases of its 2018--2023 evolution. This is supplemented by contemporaneous X-ray and radio monitoring.


\begin{table*}[ht!]
	\centering
	\caption{Spectroscopic epochs and X-ray and radio properties.}
	
	\begin{threeparttable}
	\begin{tabular}{c c c c c c c c }
		\hline
		\hline
		Epoch & MJD (date) & Instrument & X-ray\tnote{$\S$} & Radio\tnote{$\dagger$} & Br$\gamma$ EW [\AA]\tnote{$\ddagger$} & Br$\gamma$ FWHM [\kms] & $K_\mathrm{s}$ (AB mag)
		\\
		\hline
        \textbf{E--18A} & 58263.207 (2018-05-25) & GTC/EMIR & high & quiet & $10.8\pm0.7$ & $590\pm30$ & $14.53\pm0.12$ \\
        \textbf{E--18B} & 58305.005 (2018-07-06) & GTC/EMIR & high & quiet & $9.7\pm0.6$ & $630\pm30$ & -- \\
        \textbf{E--18C} & 58315.951 (2018-07-16) & GTC/EMIR & high & quiet & $6.5\pm0.9$ & $620\pm60$ & -- \\
        \textbf{E--18D} & 58354.025 (2018-08-24) & GTC/EMIR & high & quiet & $3.7\pm1.2$ & $600\pm150$ & -- \\
        \textbf{E--19A} & 58626.323 (2019-05-23) & VLT/X-shooter & low & loud & $66.9\pm0.1$ & $2090\pm20$ & $14.18\pm0.03$ \tnote{a} \\
        \textbf{E--19B} & 58633.308 (2019-05-30) & VLT/X-shooter & low & loud & $33.4\pm0.1$ & $1058\pm14$ & $\sim 14.9$ \tnote{b} \\
        \textbf{E--19C} & 58640.284 (2019-06-06) & VLT/X-shooter & low & loud & $59.6\pm0.1$ & $1740\pm30$ & $\sim 13.6$ \tnote{b} \\
        \textbf{E--19D} & 58659.199 (2019-06-25) & VLT/X-shooter & low & loud & $27.5\pm0.1$ & $813\pm11$ & $\sim 13.9$ \tnote{b} \\
        \textbf{E--21A} & 59446.019 (2021-08-20) & GTC/EMIR & flare & quiet & $7.8\pm0.5$ & $660\pm40$ & $15.33\pm0.03$ \\
		\hline
        \multirow{3}{*}{\textbf{E--21B}} & 59451.095 (2021-08-25) & \multirow{3}{*}{GTC/EMIR} & \multirow{3}{*}{flare} & \multirow{3}{*}{quiet} & \multirow{3}{*}{$6.1\pm0.4$} & \multirow{3}{*}{$486\pm13$} & \multirow{3}{*}{$15.10\pm0.05$} \\
        & 59451.989 (2021-08-25) & & & & & & \\
        & 59453.075 (2021-08-27) & & & & & & \\
		\hline
        \textbf{E--21C} & 59476.904 (2021-09-19) & GTC/EMIR & low & quiet & $8.4\pm0.7$ & $810\pm90$ & $14.82\pm0.04$ \\
		\hline
        \multirow{2}{*}{\textbf{E--22A}} & 59828.021 (2022-09-06) & \multirow{2}{*}{GTC/EMIR} & \multirow{2}{*}{low} & \multirow{2}{*}{quiet} & \multirow{2}{*}{$<0.17$} & \multirow{2}{*}{--} & \multirow{2}{*}{--} \\
        & 59830.052 (2022-09-08) & & & & & & \\
		\hline
        \textbf{E--22B} & 59859.906 (2022-10-07) & GTC/EMIR & low & quiet & $<0.17$ & -- & $15.8\pm0.2$ \\
		\hline
        \multirow{3}{*}{\textbf{E--22C}} & 59891.803 (2022-11-08) & \multirow{3}{*}{GTC/EMIR} & \multirow{3}{*}{low} & \multirow{3}{*}{quiet} & \multirow{3}{*}{$<0.17$} &  & \\
        & 59894.810 (2022-11-11) & & & & & -- & $15.7\pm0.2$ \\
        & 59895.811 (2022-11-12) & & & & & & \\
		\hline
        \textbf{E--23A} & 60076.375 (2023-05-12) & VLT/X-shooter & low & loud & $138.3\pm0.1$ & $641\pm7$ & $\sim 12.7$ \tnote{b} \\
        \textbf{E--23B} & 60082.374 (2023-05-18) & VLT/X-shooter & low & loud & $99.1\pm0.1$ & $626\pm6$ & $\sim 12.2$ \tnote{b} \\
        \textbf{E--23C} & 60094.285 (2023-05-30) & VLT/X-shooter & low & loud & $142.7\pm0.1$ & $674\pm6$ & $\sim 12.3$ \tnote{b} \\
        \textbf{E--23D} & 60161.195 (2023-08-05) & VLT/X-shooter & low & loud & $37.31\pm0.05$ & $620\pm7$ & $\sim 12.9$ \tnote{b} \\
        \hline
	\end{tabular}
    \begin{tablenotes}
        \item[$\S$] The label `high' indicates X-ray fluxes $\geq 0.2\times$ 10$^{-8}$ erg cm$^{-2}$ s$^{-1}$, whereas `low' corresponds to $< 0.1\times$ 10$^{-8}$ erg cm$^{-2}$ s$^{-1}$. Observations taken during X-ray flares (`flare') have associated fluxes in the range of $\sim$ 0.1--0.2 $\times 10^{-8}$ erg cm$^{-2}$ s$^{-1}$ (Fig. \ref{figLightCurves}).
        \item[$\dagger$] Radio-quiet periods exhibit fluxes of $\sim$ 1--5 mJy, whereas radio-loud epochs range from $\sim$ 10 to 2000 mJy (Fig. \ref{figLightCurves}). 
        \item[$\ddagger$] We define EW as positive in emission.
        \item[a] From the value $K_\mathrm{s}$ = $12.33\pm0.03$ (Vega) obtained by \citet{Vishal2019} four days before E-19A. We use $m_\mathrm{AB}-m_\mathrm{Vega}=1.85$ \citep{Blanton2007}.
        \item[b] Derived from the flux calibrated spectra and comparison with E-19A (see Sect. \ref{secKBandPhotometry}).
    \end{tablenotes}

	\label{tableEpochs}
	\end{threeparttable}
\end{table*}

\section{Observations and data reduction}
\label{secObservationsReduction}

\subsection{NIR spectroscopy}
\label{secSpectraDB}

We present NIR spectroscopy of GRS~1915+105 obtained over 18 epochs between 2018 and 2023 (see Table \ref{tableEpochs}). Eight correspond to the two radio-loud phases (2019 and 2023; four epochs each; see Fig. \ref{figLightCurves}) and these were performed with the NIR arm (10200--24750~\AA) of the X-shooter Echelle spectrograph \citep{Vernet2011} attached to the Very Large Telescope (VLT; Cerro Paranal, Chile). The remaining ten observing epochs were obtained during radio-quiet phases using the EMIR spectrograph \citep{Garzon2022} at the 10.4-m Gran Telescopio Canarias (GTC; La Palma, Spain).

Each of the VLT/X-shooter epochs consist of 16 exposures, with total exposure times ranging from 1920 to 2400~s, and a slit width of 0.9 arcsec, yielding a spectral resolution of $\sim \mathrm{54~km~s^{-1}}$. The data were processed and calibrated in wavelength and flux using the X-shooter ESO pipeline v.~3.5.0. 

For the GTC/EMIR epochs, we obtained eight exposures per observation with the K grism (20270--23730 \AA). The total exposure time per observation was \mbox{1280 s} in 2018 and \mbox{1600 s} in 2021 and 2022. We used a slit widths of 0.8 and 1.0 arcsec, which give spectral resolutions of $\sim \mathrm{82~km~s^{-1}}$ and $\sim \mathrm{103~km~s^{-1}}$, respectively. The data were reduced using the EMIR Data Reduction Pipeline PyEMIR v.~0.15 \citep{Pascual2010,Cardiel2019}. The wavelength calibration was carried out using the OH sky lines and verified with the arc lamp files.

\begin{figure*}[ht]
    \centering
    \includegraphics[width=12truecm,height=8truecm]{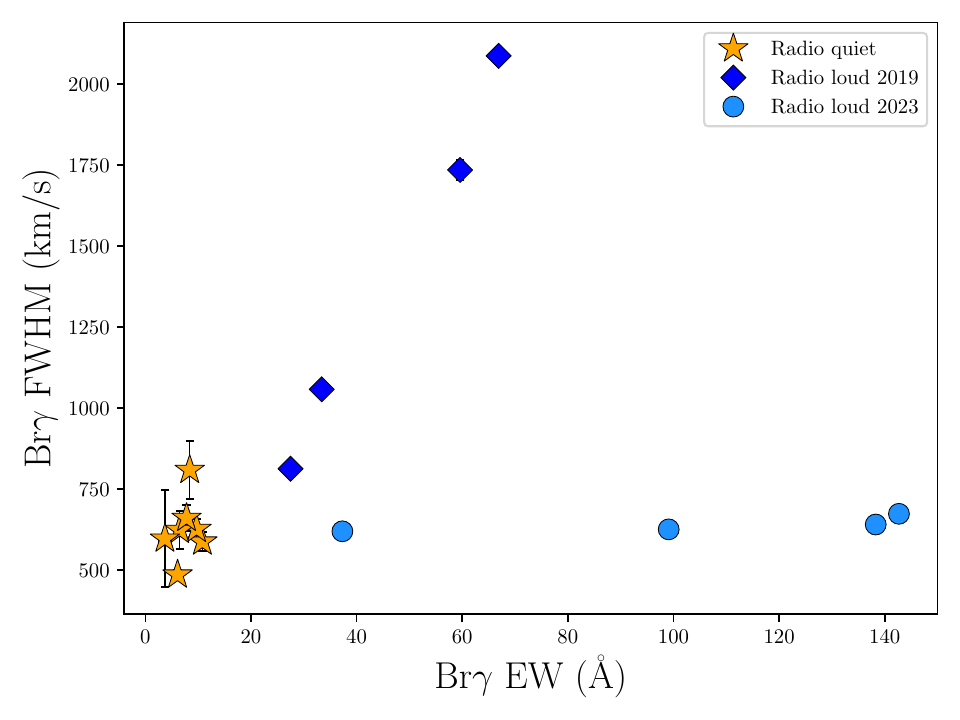}
    \includegraphics[width=17truecm,height=12truecm]{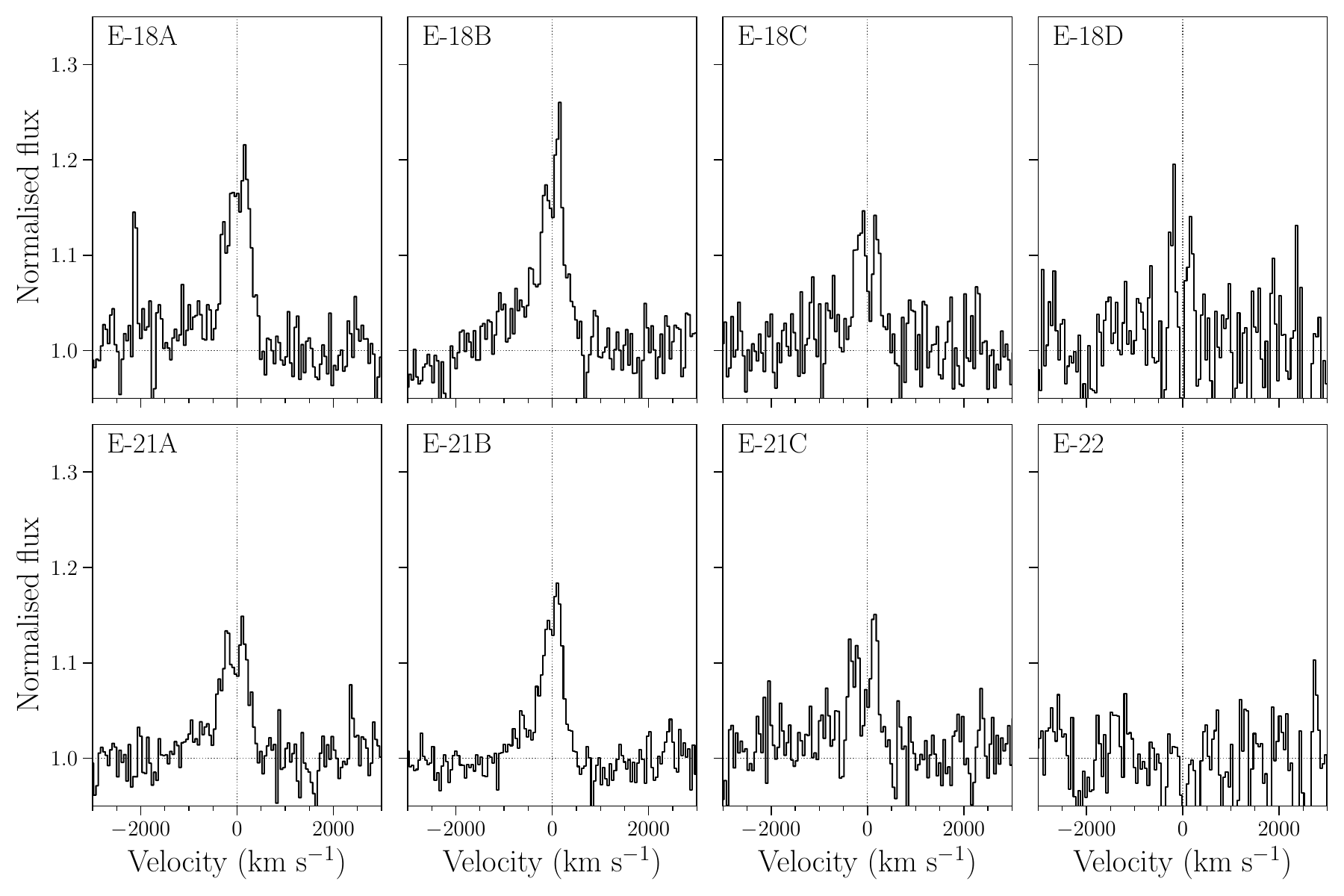}
    \caption{Evolution of the Br$\gamma$ emission line. \textit{Top panel:} EW vs FWHM for Br$\gamma$ line. \textit{Bottom panels:} evolution of the Br$\gamma$ line profile as observed by GTC/EMIR. We note that E-22 correspond to the mean of the data from the three 2022 epochs.}
    \label{figEMIREpochs}
\end{figure*}

The VLT and GTC spectra were corrected for telluric absorption lines using \textsc{Molecfit} v. 1.5.9 and v. 3.0.3 \citep{Smette2015} following the method described in \citet{Sanchez-Sierras2023}. The data analysis was performed using \textsc{Molly} and custom routines (\textsc{Python} 3.7). Table \ref{tableEpochs} includes the equivalent width (EW) and the full-width at half maximum (FWHM) of Br$\gamma$ emission line (21655 \AA) for every epoch. The latter was measured from a Gaussian fit to the line profile.

In order to increase the signal-to-noise ratio of the GTC/EMIR spectra, we grouped observations obtained close in time into single epochs. Prior to this, we checked that the Br$\gamma$ line profiles were consistent within each epoch. Thus, E-22A includes two observations, while E-21B and E-22C contain three observations each (Table \ref{tableEpochs}). On the other hand, the VLT/X-shooter epochs correspond to a single observing night. In an attempt to search for possible short-term variability, we also tried dividing each 40-min epoch into eight 5-min spectra, but we did not find any clear evidence for it (e.g. see Fig. \ref{figE19CHighTimeRes} for epoch E-19C). Therefore, only the average spectrum of each GTC and VLT epoch is discussed hereafter (see Table \ref{tableEpochs}).

\subsection{K-band photometry}
\label{secKBandPhotometry}

Table \ref{tableEpochs} includes the $K_\mathrm{s}$ magnitude of the system for each spectroscopic epoch. Those corresponding to GTC observations were obtained, when possible, from the EMIR acquisition images.

The X-shooter acquisition is performed in the optical bands. In order to estimate the NIR flux, we have taken the following steps. First, we assigned the $K_\mathrm{s}$ AB magnitude ($14.18\pm0.03$) reported in \citet{Vishal2019} to E-19A. This was obtained only four days before E-19A. Second, we multiplied each flux-calibrated spectrum by the $K_\mathrm{s}$ filter transmission curve and integrated the resulting flux. Finally, we determined the $K_\mathrm{s}$ magnitude for each epoch, using that of E-19A as a reference. This method might be affected by systematic errors, as it relies on the spectral flux calibration. Nevertheless, the resulting magnitudes indicate that the radio-loud events are associated with higher NIR fluxes (see Table \ref{tableEpochs}). Furthermore, it shows that the 2023 radio-loud event (Fig. \ref{figLightCurves}) was brighter both in radio and in the NIR.

\subsection{X-ray and radio monitoring}
We extracted the 2018--2023 X-ray light curve in the 2--12~keV interval from data and procedures\footnote{http://maxi.riken.jp/mxondem/} provided by MAXI \citep{Matsuoka2009}. The resulting light curve with one day bin size is shown in Fig. \ref{figLightCurves}.

We also used radio data obtained with the AMI during 2018--2020 (\mbox{15.5 GHz}; \citealt{Zwart2008}) and MeerKAT (\mbox{1.28 GHz}; \citealt{Fender2016a}) arrays, as presented in \citet{Motta2021}. These have been supplemented by new AMI observations covering the evolution of the system until August 2023. The data have been processed and reduced in the same way as described in \citet{Motta2021}. Table \ref{tableEpochs} labels the X-ray (high, flare and low) and radio (loud, quiet) behaviour at the time of each spectroscopic epoch (based on Fig. \ref{figLightCurves}).

\section{Analysis and results}
\label{SecResults}

The 18 spectroscopic epochs presented in this work cover the evolution of GRS~1915+105 from May 2018 to August 2023 (Table \ref{tableEpochs}). During this time, we identified four different stages of radio and X-ray activity, which are also associated with significant changes in the NIR spectrum. We discuss them in turn below.

\subsection{X-ray decay}

After more than 25 years of intense X-ray activity, GRS~1915+105 entered a new phase characterised by a gradual decrease of the X-ray flux, reaching a low-luminosity plateau in 2018 (MJD 58350; Fig. \ref{figLightCurves}). Epochs E-18A and E-18B (GTC), still taken during a phase of relatively high X-ray activity, are characterised by a fairly strong asymmetric Br$\gamma$ line in emission (EW $\sim$ 10 \AA; see Table \ref{tableEpochs}). The line also shows a hint of a narrow double-peaked profile. The EW of Br$\gamma$ decreased in E-18C and E-18D (down to $\sim 6.5$ and 4 \AA, respectively), also developing a more pronounced double-peak (Table \ref{tableEpochs} and Fig. \ref{figEMIREpochs}). These profiles are commonly interpreted as signalling the presence of an accretion disc (\citealt{Smak1969}). In this context, the evolution from E-18A to E-18D is consistent with the outer accretion disc responding to the observed drop in X-rays. Thus, as the irradiating X-ray emission decreases, the line-forming region shifts towards inner, higher-velocity regions of the disc, producing less-intense profiles and larger double-peak separations. This behaviour is commonly observed in optical and NIR emission lines of XRB transients when approaching quiescence (e.g. \citealt{Casares2015}).

\subsection{2019--2021 radio-loud phase}

\begin{figure*}[ht]
    \centering
    \includegraphics[width=12truecm]{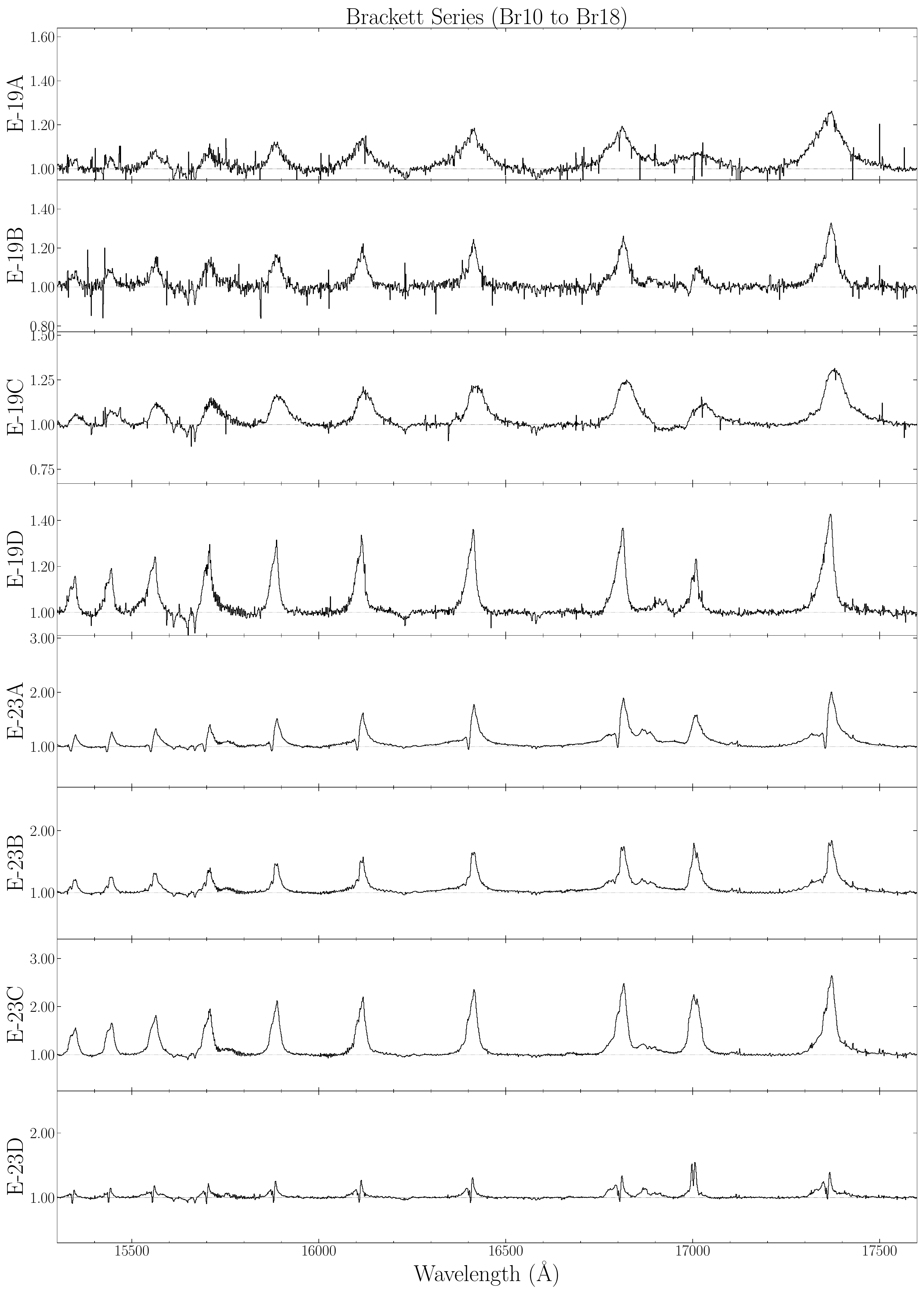}
    \includegraphics[width=3truecm]{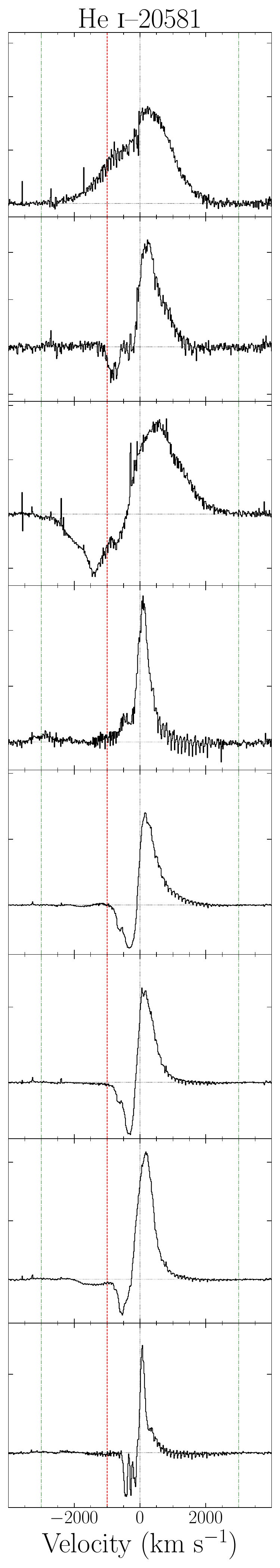}
    \includegraphics[width=3truecm]{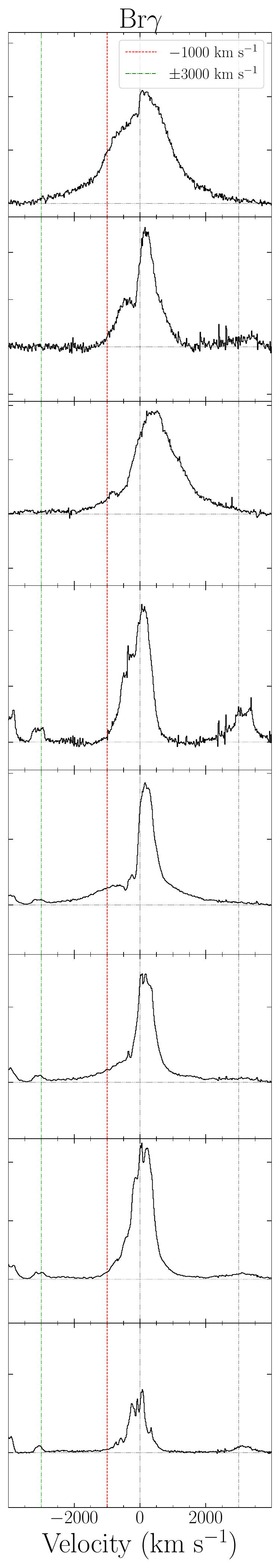}
    \caption{High order transitions of Brackett Series, \hei--20581 and Br$\gamma$ during the eight epochs observed by VLT/X-shooter. We note that vertical scales are different for every epoch. Red (dashed) and green (dash-dotted) vertical lines indicate velocities of $-1000$ \kms and $\pm \mathrm{3000~km~s^{-1}}$, respectively.}
    \label{figVLT8Epochs}
\end{figure*}

The next significant change was observed in May 2019 (MJD$\sim$58600). This involved a further dimming of the X-ray emission, but this time it was accompanied by an increase in the radio flux (Fig. \ref{figLightCurves}), reaching pre-2018 levels \citep{Motta2021}, and in the NIR emission (from $K_\mathrm{s}$ $\sim14.5$ to 13.6, see Table \ref{tableEpochs}). Four VLT/X-shooter epochs were taken during the first month of this radio-loud phase (E-19A to E-19D). These spectra are very rich in emission lines, from \ion{He}{i}--10830 to Br$\gamma$ (21655 \AA), with highly variable properties (Fig. \ref{figVLT8Epochs}). Given the high quality spectra and the accurate telluric correction, we even detect the Pa$\alpha$ line and high-order Pfund transitions in emission (middle and bottom panels in Figs. \ref{figfullNIR19A}-\ref{figfullNIR19D}). These features are usually difficult to detect from ground-based observatories since they are affected by strong telluric bands.

\begin{figure*}[ht]
    \centering
    \includegraphics[width=8.5truecm]{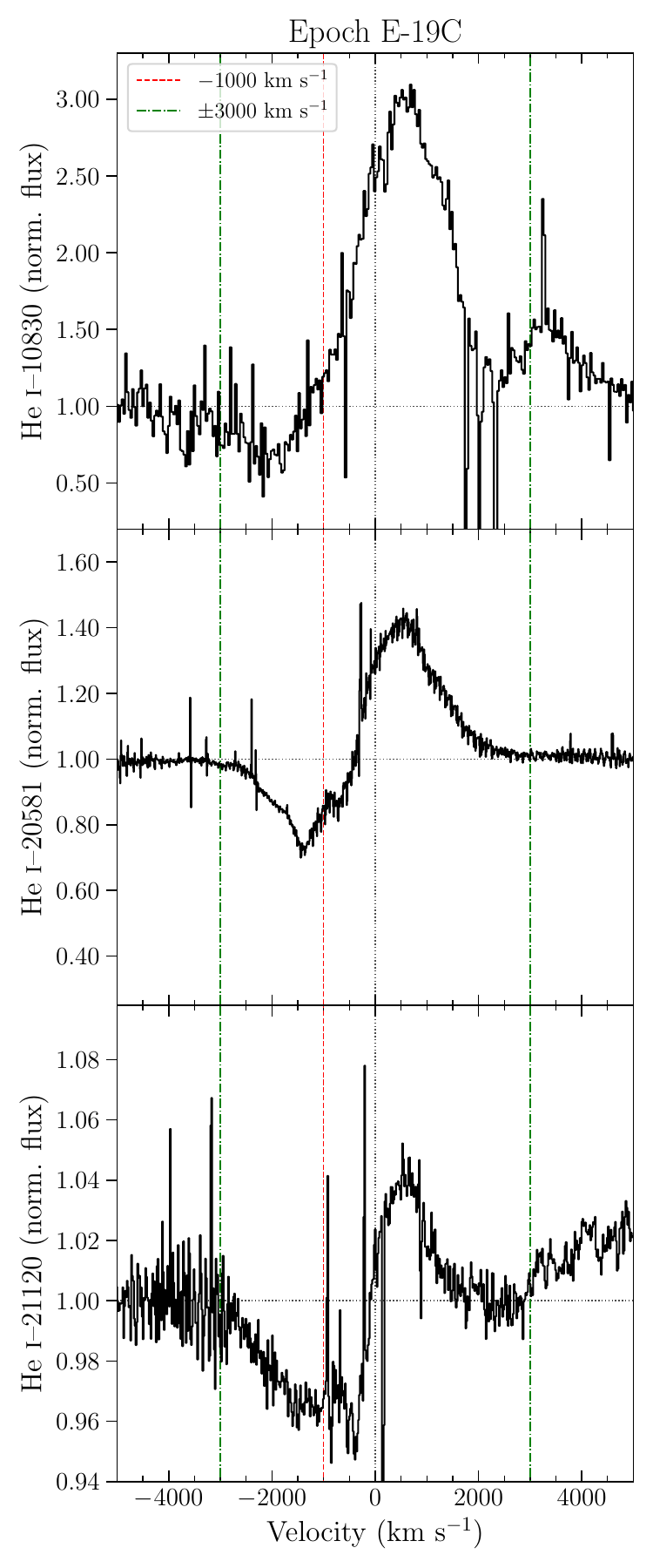}
    \includegraphics[width=8.5truecm]{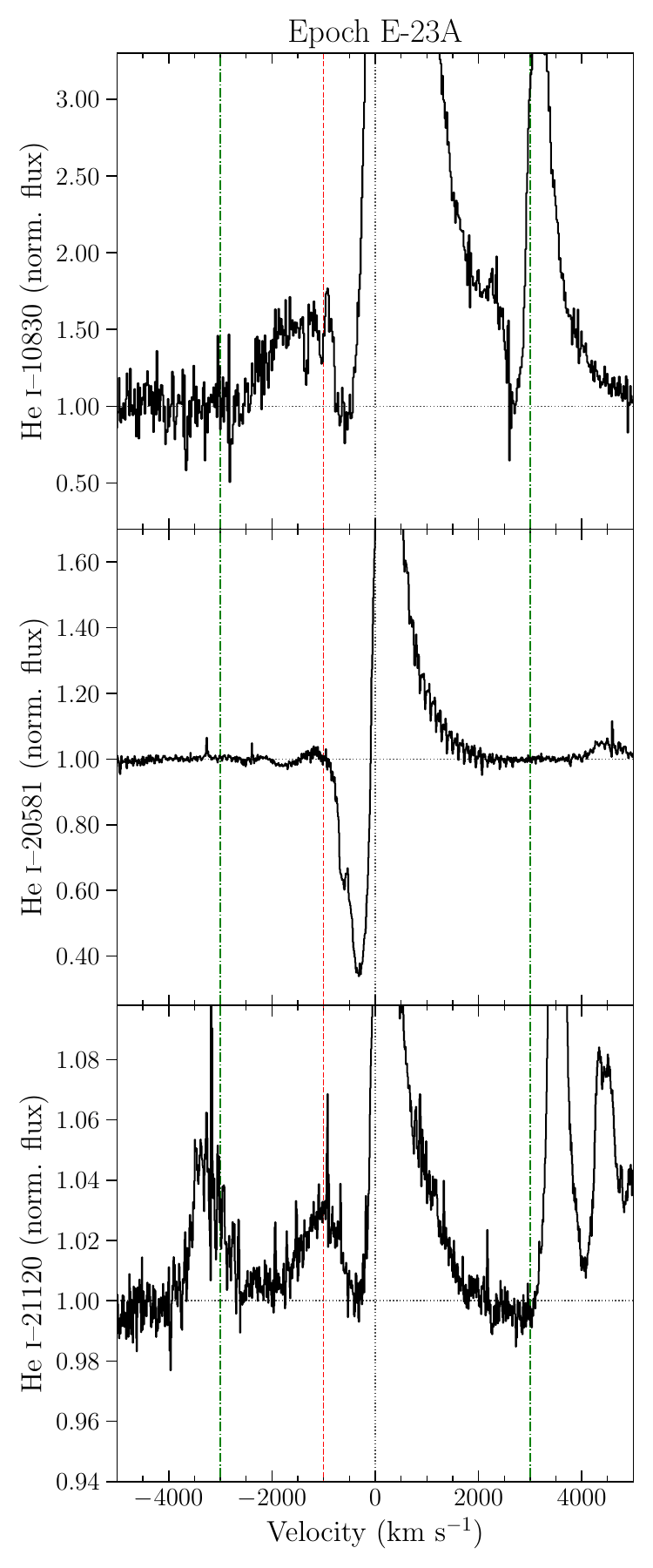}
    \caption{\ion{He}{i} line profiles (10830~\AA, 20581~\AA~and 21120~\AA) during epochs E-19C (left) and E-23A (right). Vertical lines are the same as in Fig. \ref{figVLT8Epochs}.}
    \label{figHeIWinds}
\end{figure*}

Compared to E-18A-D, the EW of Br$\gamma$ increases dramatically to $\sim 67$~\AA~in E-19A, becoming single-peaked but with very broad wings that extend to a full width at zero intensity of $\sim~\mathrm{6000~km~s^{-1}}$. This behaviour is also seen in other Brackett lines and \ion{He}{i} transitions (see Fig. \ref{figVLT8Epochs}). Even more striking is the evidence of conspicuous P-Cygni line profiles only one week later (epochs E-19B and E-19C), indicating the presence of a wind-type outflow (e.g. \citealt{Castor1979}). These profiles are detected in the main \hei\ lines, reaching terminal velocities (hereafter defined as the blue-edge of the blue-shifted absorption) of $\sim \mathrm{3000~km~s^{-1}}$. The left panels in Fig. \ref{figHeIWinds} show examples of these features (see also Fig. \ref{figVLT8Epochs}). The deepest P-Cygni profile is seen in \ion{He}{i}--20581, which shows a blue-shifted absorption reaching $\sim 30\%$ below the continuum level. In addition, we also detect P-Cygni-like features in \ion{He}{i}--10830, and \ion{He}{i}--21120, all of them with similar terminal velocities. No obvious P-Cygni profiles were detected during E-19D, although the emission lines remain strong and have complex profiles (e.g. skewed towards the red). In addition, two emission lines are detected in E-19D at $\sim 21370$ and $\sim 21435$~\AA,~which we identify as \ion{Mg}{ii} transitions. These transitions have been previously seen in hot stars such as early-B supergiants \citep{Hanson2005}. These are also weakly detected in epochs E-23A-D.

\subsection{2021--2023 radio-quiet phase}

The system returned to lower radio flux levels (i.e. similar to those of early 2019) in May 2021, and remained in this regime until April 2023 (\mbox{$\sim$ MJD 59350;} Fig. \ref{figLightCurves}). In turn, the X-ray flux remained at the lowest values most of the time. The notable exceptions to this are the X-ray flares detected around MJD 59100 and 59400 (see Fig. \ref{figLightCurves}; \citealt{Motta2021}). In particular, the latter flare occurred at the beginning of this radio-quiet period and was covered by three GTC/EMIR epochs: \mbox{E-21A} and E-21B during the flare, while E-21C just after the event. The first two (particularly E-21B) show a Br$\gamma$ emission line profile very reminiscent of that of E-18A-B (i.e. still during the X-ray active phase), albeit less intense (Fig. \ref{figEMIREpochs}). By contrast, E-21C is very similar to E-18C-D, with a pronounced double-peaked line profile (see Fig. \ref{figEMIREpochs} and Table \ref{tableEpochs}). This evolution is consistent with the outer accretion disc (traced by the Br$\gamma$ emission) responding to the changes in the observed X-ray flux. 

We took three additional GTC/EMIR epochs (E-22A to \mbox{E-22C}) during late 2022. They all show a featureless spectrum, from which we determine a $3\sigma$ upper limit to the Br$\gamma$ EW of 0.17~\AA~(see averaged spectrum in Fig. \ref{figEMIREpochs}). GRS~1915+105 reached the faintest NIR emission during these epochs with $K_\mathrm{s}$~$\sim15$.

\subsection{2023 radio-loud phase}
\label{sec2023radioloud}

A new episode of enhanced radio emission at low X-ray flux started in April 2023 \citep{Egron2023}. Four VLT/X-shooter spectra were obtained during this phase (May to August 2023). These are qualitatively similar to those from 2019 (see Figs. \ref{figfullNIR23A} to \ref{figfullNIR23D}). First, the P-Cygni line profiles are detected in \ion{He}{i} transitions (see Fig. \ref{figHeIWinds}, right panel) with blue-shifted absorptions reaching depths below $\sim 60\%$ of the continuum level in some cases (e.g. E-23A; see Fig. \ref{figHeIWinds}, middle-right panel). The terminal velocities are $\sim$ $\mathrm{1000~km~s^{-1}}$ ($\sim \mathrm{750~km~s^{-1}}$ in E-23D), significantly lower than in 2019. Second, broad emission line wings are detected in the Brackett series (Fig. \ref{figVLT8Epochs}, bottom panels), which also show blue absorption troughs during E-23A and E-23D. The emission features are also narrower than in 2019 (see top panel of Fig. \ref{figEMIREpochs}). However, in E-23A and E-23B, there are additional emission components that are broader (see Br$\gamma$ in Fig. \ref{figVLT8Epochs}), meeting the continuum at $\sim \mathrm{3000~km~s^{-1}}$. Finally, it is worth mentioning the particularly complex line profiles observed in E-23D, with additional emission peaks (e.g. Br$\gamma$), noting that some of them found within the blue-shifted absorption of the P-Cygni profiles (e.g. \ion{He}{i}--20581). Interestingly, this last spectrum is contemporaneous with a weak X-ray flare (Fig. \ref{figLightCurves}).

\section{Discussion}
\label{secDiscussion}

\subsection{NIR outflows during the radio-loud phases}

The VLT/X-shooter epochs (2019 and 2023), taken during radio-loud phases at low X-ray flux, exhibit wind-type outflows. This is evident from the presence of strong P-Cygni profiles, particularly in the NIR \ion{He}{i} transitions. Optical P-Cygni profiles with comparably deep blue-shifted absorptions (i.e.~$\mbox{>30}$\% below the continuum level) have also been observed in V404~Cyg and V4641~Sgr \citep{Munoz-Darias2016, Munoz-Darias2017, Munoz-Darias2018}. Together with GRS 1915+105, these systems have the longest orbital periods and, therefore, the largest accretion discs among BH transients. Furthermore, they exhibited the fastest cold (optical/infrared) winds among XRBs ($\mathrm{\sim 3000~km~s^{-1}}$; see e.g. \citealt{Panizo-Espinar2022}). In addition, V404~Cyg and V4641~Sgr displayed high levels of intrinsic X-ray absorption (e.g. \citealt{Morningstar2014, Motta2017a, Koljonen2020}), as is the case for GRS~1915+105 during the radio-loud phases ($N_\mathrm{H} \sim 7 \times 10^{23}$ cm$^{-2}$; \citealt{Balakrishnan2021, Miller2020,Motta2021}). 

The NIR signatures of accretion disc winds have been detected in other BH transients, such as MAXI~J1820+070 \citep{Sanchez-Sierras2020}, MAXI~J1348-630 \citep{Panizo-Espinar2022}, and MAXI J1803-298 \citep{MataSanchez2022}. However, the NIR P-Cygni profiles reported in this work are the most prominent seen in a BH transient to date. The observational appearance of the wind in GRS~1915+105 differs between 2019 and 2023. In 2019, we measure larger blue-edge velocities in the P-Cygni profiles, reaching $\mathrm{\sim 3000~km~s^{-1}}$, whereas in 2023 some emission lines (e.g. Brackett series) displayed additional emission components with blue wings meeting the continuum at $\mathrm{\sim 3000~km~s^{-1}}$. This combination of blue-shifted absorptions and enhanced emission at the same velocities has been observed in other systems and may be related to the optical depth of the ejecta (e.g. Fig. 15 in \citealt{MataSanchez2018}). Furthermore, the evolution of the EW and FWHM of the Br$\gamma$ line (top panel in Fig. \ref{figEMIREpochs}) is significantly different between 2019 and 2023. Comparing this behaviour with that of H$\alpha$ during the 2015 outburst of V404 Cyg (Fig. 9 in \citealt{MataSanchez2018}), we are able to identify the 2019 spectra with the phase of strongest P-Cygni signatures, while the evolution during 2023 might be characterised by optically thinner ejecta. Finally, in 2023, we observe increased radio brightness and more intense Br$\gamma$ lines, as compared to 2019, resembling the early outburst behaviour noted by \citet{Eikenberry1998}.

\subsection{A massive, multi-phase wind}
In this work, we have discovered strong NIR outflow signatures in GRS~1915+105 at low X-ray fluxes. However, weaker (tentative) NIR detections had previously been reported at much higher X-ray fluxes (e.g. \citealt{Mirabel1997, Marti2000}). Conversely, hot winds at similar velocities than those reported here have been widely observed in the source (e.g. \citealt{Kotani2000,Lee2002, Ueda2009, Neilsen2011}) during the (X-ray) bright phases. 

X-ray accretion disc winds were convincingly detected during the onset of the 2019 radio-loud event (\citealt{Miller2020}), 24 days before our first NIR wind detection \mbox{(E-19A)}. However, the characteristic velocity of this hot wind was $\mathrm{\sim 300~km~s^{-1}}$, which is an order of magnitude lower than that of the NIR wind and previous X-ray detections. This is consistent with the idea that accretion disc winds in XRBs are multi-phase in nature, with a variable balance between the different phases (e.g. hot and cold) depending on the physical conditions of the ejecta. This scenario is supported by the observation of both X-ray and optical winds in V404 Cyg \citep{Munoz-Darias2022}. Thus, contemporaneous observations at low and high energies are likely the most effective approach for gaining a deeper understanding of the observational properties of the wind. In this regard, a mass outflow rate comparable to the accretion rate was roughly estimated for the low-velocity X-ray wind (\citealt{Miller2020}); however, this could be increased by an order of magnitude if the NIR wind velocity were to be used instead. This would suggest a massive, outflowing wind, which is consistent with previous estimations for this system (e.g. \citealt{Neilsen2011}) and V404 Cyg (\citealt{Munoz-Darias2016, Casares2019}).

\subsection{Long-term evolution of GRS~1915+105 and the nature of the radio-loud events}

The evolution of the Br$\gamma$ line during the radio-quiet epochs indicates that the properties of the line respond to the observed X-ray flux. For instance, we observe the line weakening and the peak-to-peak separation of the profile increasing as the system becomes fainter in X-rays, a behaviour typically observed in XRB transients when approaching quiescence. This is consistent with the X-ray fitting by \citet{Motta2021}, which shows no indication of enhanced intrinsic absorption during the radio-quiet phase. The NIR continuum also follows this trend, with the second radio-quiet period (i.e.$~2022$) being fainter in the NIR than the first one ($\sim 2019$). This makes even more significant the lack of Br$\gamma$ emission in E-22A to E-22C, since the diluting continuum is also weaker. Featureless NIR spectra have been seen in BH hard states at very low luminosity (e.g. \citealt{Sanchez-Sierras2020}) and might indicate a dominant jet contribution. All things considered, if the radio-loud epochs are excluded, the behaviour of GRS~1915+105 since 2018 is in line with a gradual decline from its long and luminous outburst, potentially heading towards quiescent levels. However, we must consider how the radio-loud epochs fit in this scenario.

A first-order explanation for the radio-loud phases is that they are associated with a severe increase of the accretion rate. The source would remain faint in X-rays as a result of very high intrinsic absorption (\citealt{Miller2020,Motta2021}). In this scenario, the radio-loud phases may correspond to mini-outbursts, also known as secondary outbursts or re-brightenings, which are commonly observed in XRB transients during the decay phase of a main outburst (e.g. \citealt{Callanan1995,Jonker2012,Cuneo2020a}). These secondary events tend to be progressively shorter and fainter than the main event (e.g. $\sim 250$~day-long outburst and 25--30~day-long reflares in MAXI J1820+070; \citealt{Stiele2020}). In the case of GRS~1915+105, the first radio-loud event lasted two years, much less than the 25-year main outburst. However, the high associated radio fluxes (record breaking in 2023; see \citealt{Trushkin2023}) may challenge this scenario if they are tracing the actual X-ray (i.e. accretion) luminosity. Nevertheless, it is important to consider that GRS~1915+105 is a unique system in many respects and the behaviour of its large accretion disc may deviate from the norm also during this potentially last phase of the outburst.

\section{Conclusions}
\label{secConclusions}

In this work, we present NIR spectroscopy covering the evolution of GRS~1915+105 from the start of its X-ray decline in 2018. Our results reveal the presence of strong NIR winds during two radio-loud phases displayed by the system, with terminal velocities of up to $\mathrm{\sim 3000~km~s^{-1}}$. These outflows are contemporaneous with high intrinsic X-ray absorption. A similar behaviour has been observed in other BH transients, in particular, V404 Cyg and V4641 Sgr. This reinforces the idea that massive and multi-phase outflows, which are potentially able to obscure the source in X-rays, are a distinctive feature of the largest BH accretion discs. In addition, the evolution of the Br$\gamma$ emission line during radio-quiet phases is consistent with the low X-ray flux displayed by the source. This study highlights the importance of NIR spectroscopy in the study of BH transients and should be complemented with future investigations on the evolution of this critical system across various wavelengths.

\begin{acknowledgements}
We are thankful to the anonymous referee for helpful comments that improved the paper. This work is supported by the Spanish Ministry of Science via an \textit{Europa Excelencia} grant (EUR2021-122010) and the \textit{Plan de Generacion de conocimiento}: PID2020-120323GB-I00 and PID2021-124879NB-I00. JAFO acknowledges financial support by the Spanish Ministry of Science and Innovation (MCIN/AEI/10.13039/501100011033) and ``ERDF A way of making Europe'' though the grant PID2021-124918NB-C44; MCIN and the European Union -- NextGenerationEU through the Recovery and Resilience Facility project ICTS-MRR-2021-03-CEFCA. We are thankful to the GTC staff, in particular Antonio L. Cabrera Lavers, for their useful help to perform the spectroscopic observations presented in this letter. Based on observations carried out with the EMIR spectrograph under programmes GTC38-18A and GTC55-21A. Based on data from the GTC Archive at CAB (INTA-CSIC). Based on observations collected at the European Southern Observatory under ESO programmes 103.200K.001 and 111.2649.001. We thank the staff of the Mullard Radio Astronomy Observatory, University of Cambridge, for their support in the maintenance and operation of AMI, which is supported by the Universities of Cambridge and Oxford. We also acknowledge support from the European Research Council under grant ERC-2012-StG-307215 LODESTONE. We thank the staff at the South African Radio Astronomy Observatory (SARAO) for scheduling the MeerKAT observations. The MeerKAT telescope is operated by the South African Radio Astronomy Observatory, which is a facility of the National Research Foundation, an agency of the Department of Science and Innovation. This research has made use of MAXI data provided by RIKEN, JAXA and the MAXI team. \textsc{Molly} software developed by Tom Marsh is gratefully acknowledged. 

\end{acknowledgements}

%
%
\bibliographystyle{aa}
\bibliography{Libreria}

\begin{appendix}

\section{Supplementary figures}

\begin{figure*}[ht]
    \centering
    \includegraphics[width=16truecm]{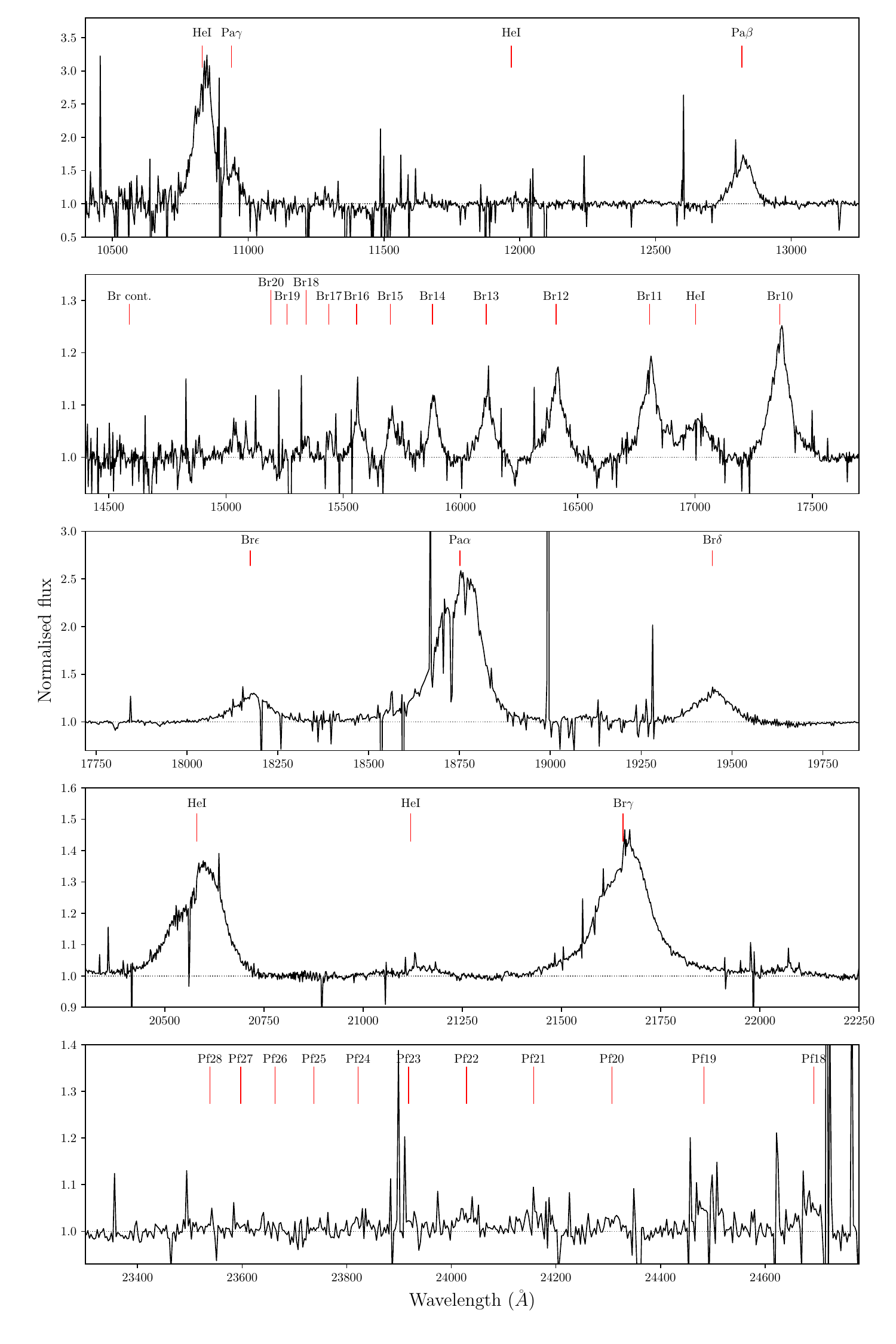}
    \caption{Full normalised spectrum of VLT/X-shooter epoch E-19A. \textit{Middle panel}: Pa$\alpha$ in the center alongside with Br$\delta$ and Br$\epsilon$. \textit{Bottom panel}: Higher orders of Pfund Series. We note that middle and bottom panels are strong affected by telluric absorptions.}
    \label{figfullNIR19A}
\end{figure*}

\begin{figure*}[ht]
    \centering
    \includegraphics[width=16truecm]{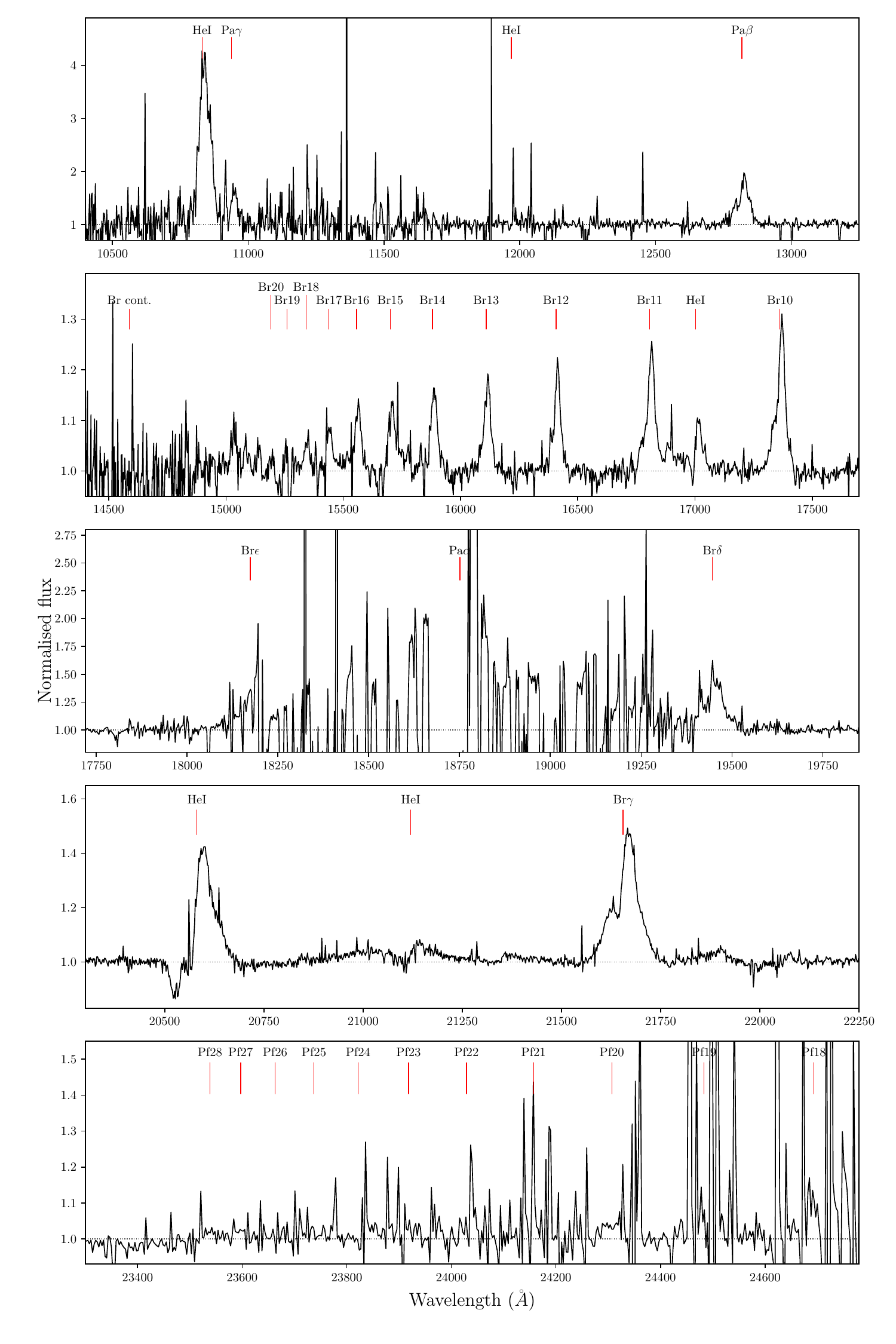}
    \caption{Full normalised spectrum of VLT/X-shooter epoch E-19B. \textit{Middle panel}: Pa$\alpha$ in the center alongside with Br$\delta$ and Br$\epsilon$. \textit{Bottom panel}: Higher orders of Pfund Series.}
    \label{figfullNIR19B}
\end{figure*}

\begin{figure*}[ht]
    \centering
    \includegraphics[width=16truecm]{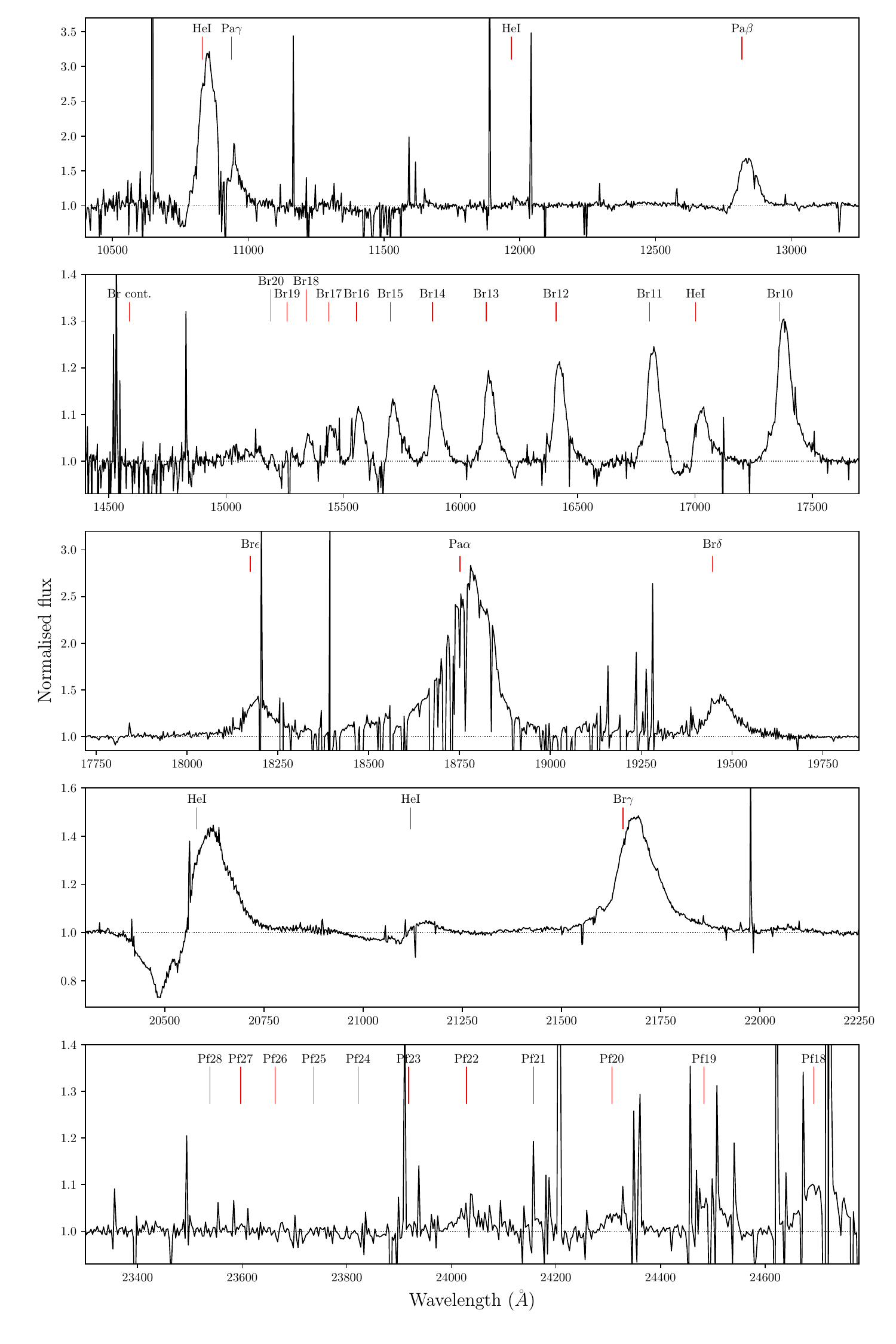}
    \caption{Full normalised spectrum of VLT/X-shooter epoch E-19C. \textit{Middle panel}: Pa$\alpha$ in the center alongside with Br$\delta$ and Br$\epsilon$. \textit{Bottom panel}: Higher orders of Pfund Series.}
    \label{figfullNIR19C}
\end{figure*}

\begin{figure*}[ht]
    \centering
    \includegraphics[width=16truecm]{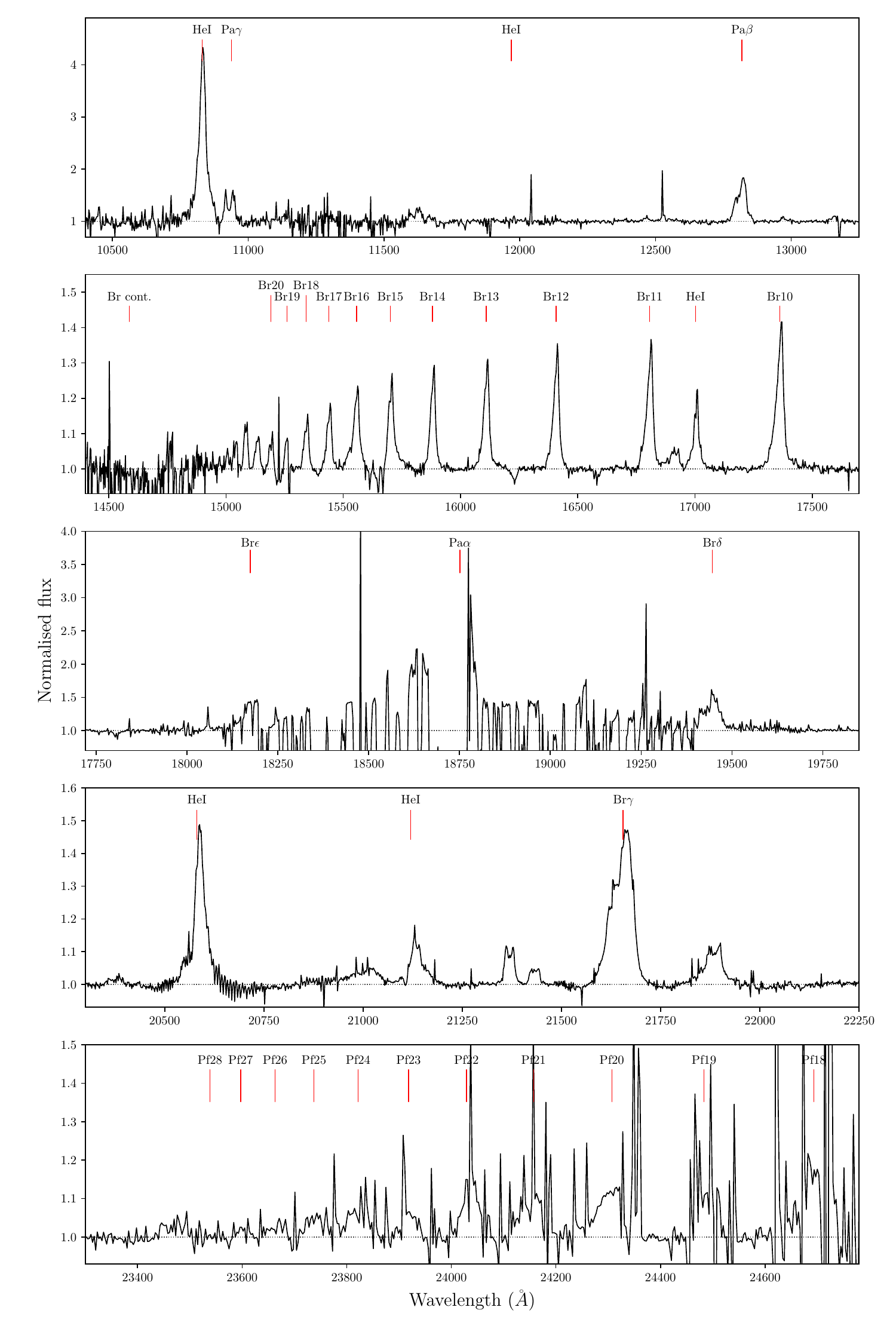}
    \caption{Full normalised spectrum of VLT/X-shooter epoch E-19D. \textit{Middle panel}: Pa$\alpha$ in the center alongside with Br$\delta$ and Br$\epsilon$. \textit{Bottom panel}: Higher orders of Pfund Series.}
    \label{figfullNIR19D}
\end{figure*}

\begin{figure*}[ht]
    \centering
    \includegraphics[width=16truecm]{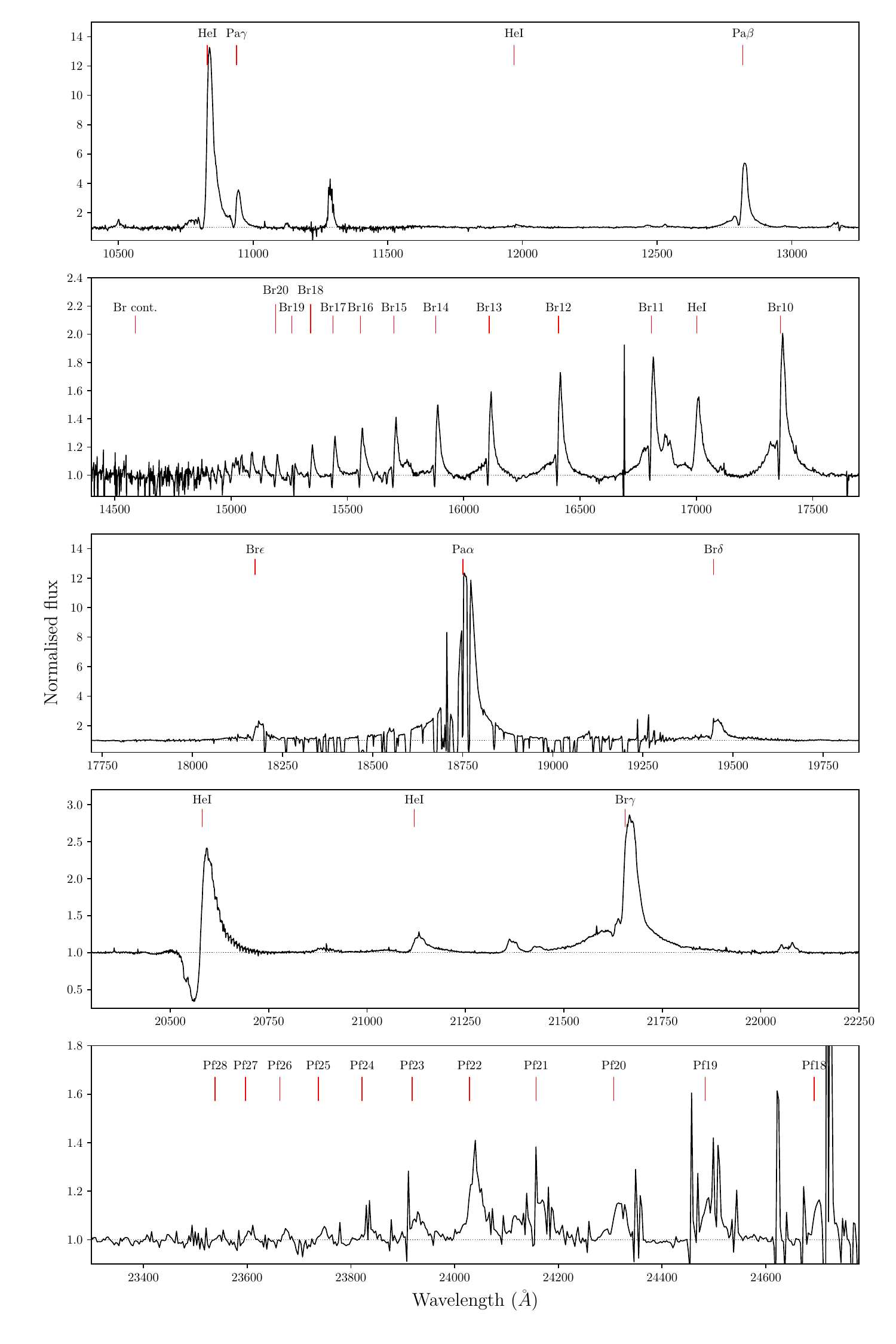}
    \caption{Full normalised spectrum of VLT/X-shooter epoch E-23A. \textit{Middle panel}: Pa$\alpha$ in the center alongside with Br$\delta$ and Br$\epsilon$. \textit{Bottom panel}: Higher orders of Pfund Series.}
    \label{figfullNIR23A}
\end{figure*}

\begin{figure*}[ht]
    \centering
    \includegraphics[width=16truecm]{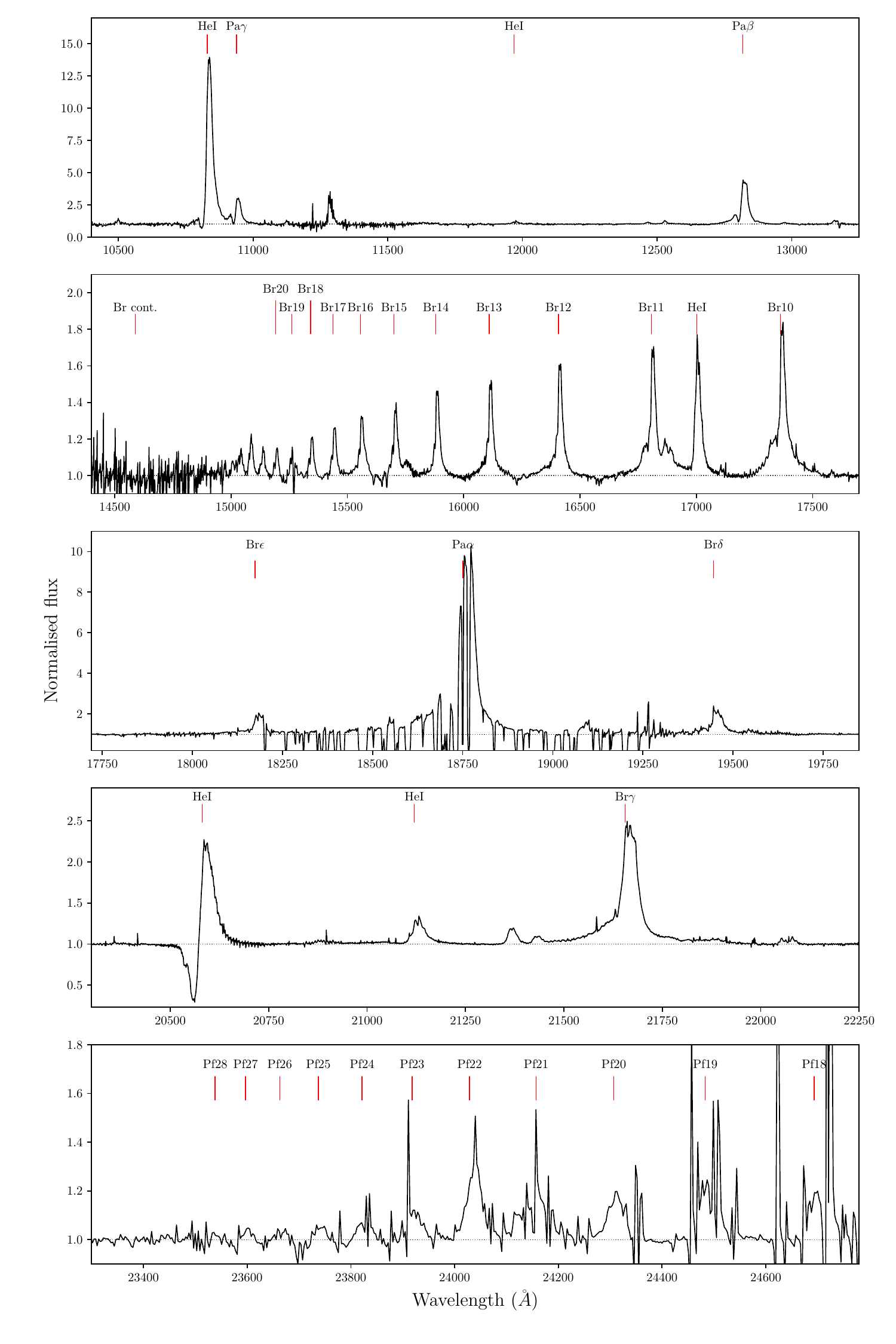}
    \caption{Full normalised spectrum of VLT/X-shooter epoch E-23B. \textit{Middle panel}: Pa$\alpha$ in the center alongside with Br$\delta$ and Br$\epsilon$. \textit{Bottom panel}: Higher orders of Pfund Series.}
    \label{figfullNIR23B}
\end{figure*}

\begin{figure*}[ht]
    \centering
    \includegraphics[width=16truecm]{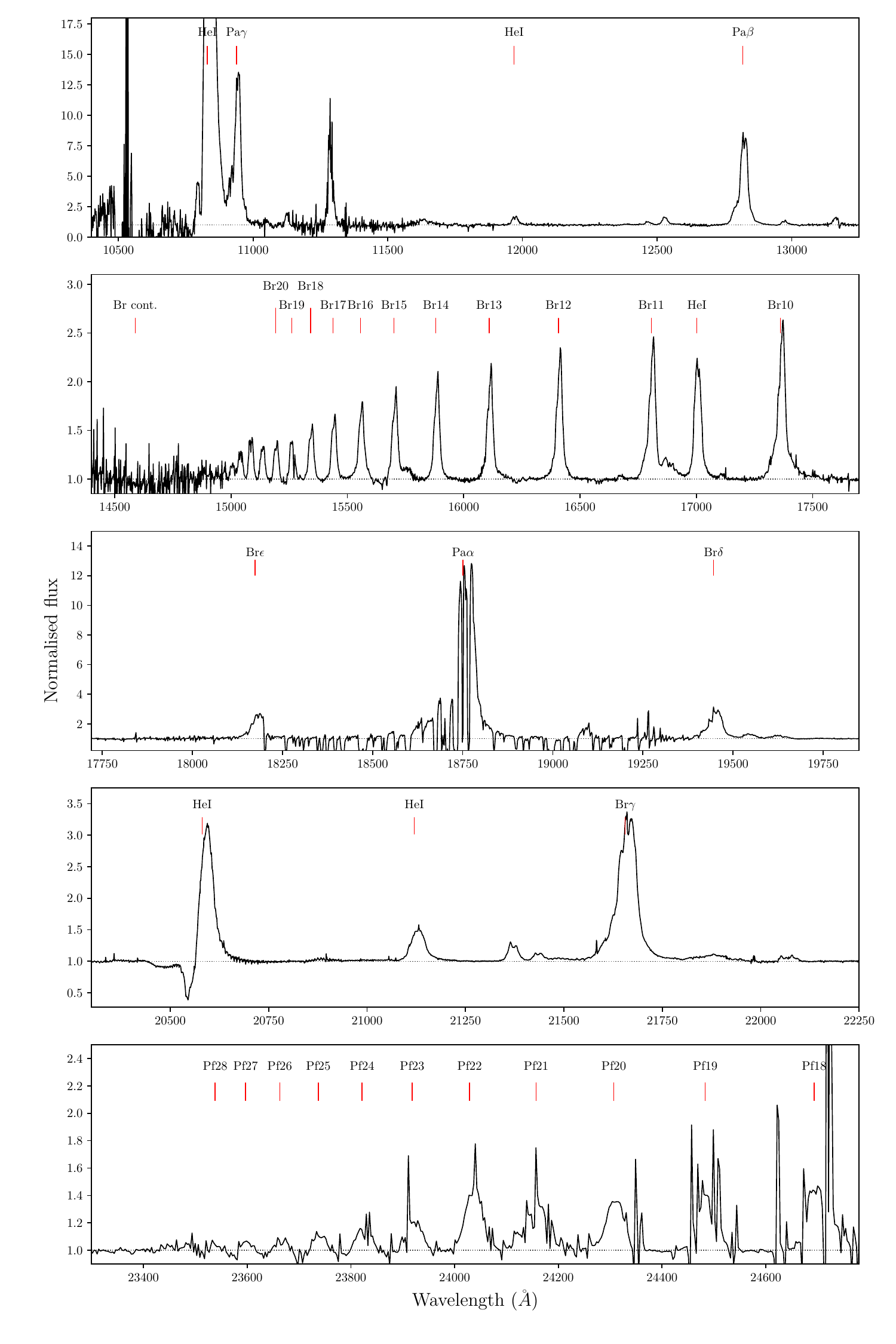}
    \caption{Full normalised spectrum of VLT/X-shooter epoch E-23C. \textit{Middle panel}: Pa$\alpha$ in the center alongside with Br$\delta$ and Br$\epsilon$. \textit{Bottom panel}: Higher orders of Pfund Series.}
    \label{figfullNIR23C}
\end{figure*}

\begin{figure*}[ht]
    \centering
    \includegraphics[width=16truecm]{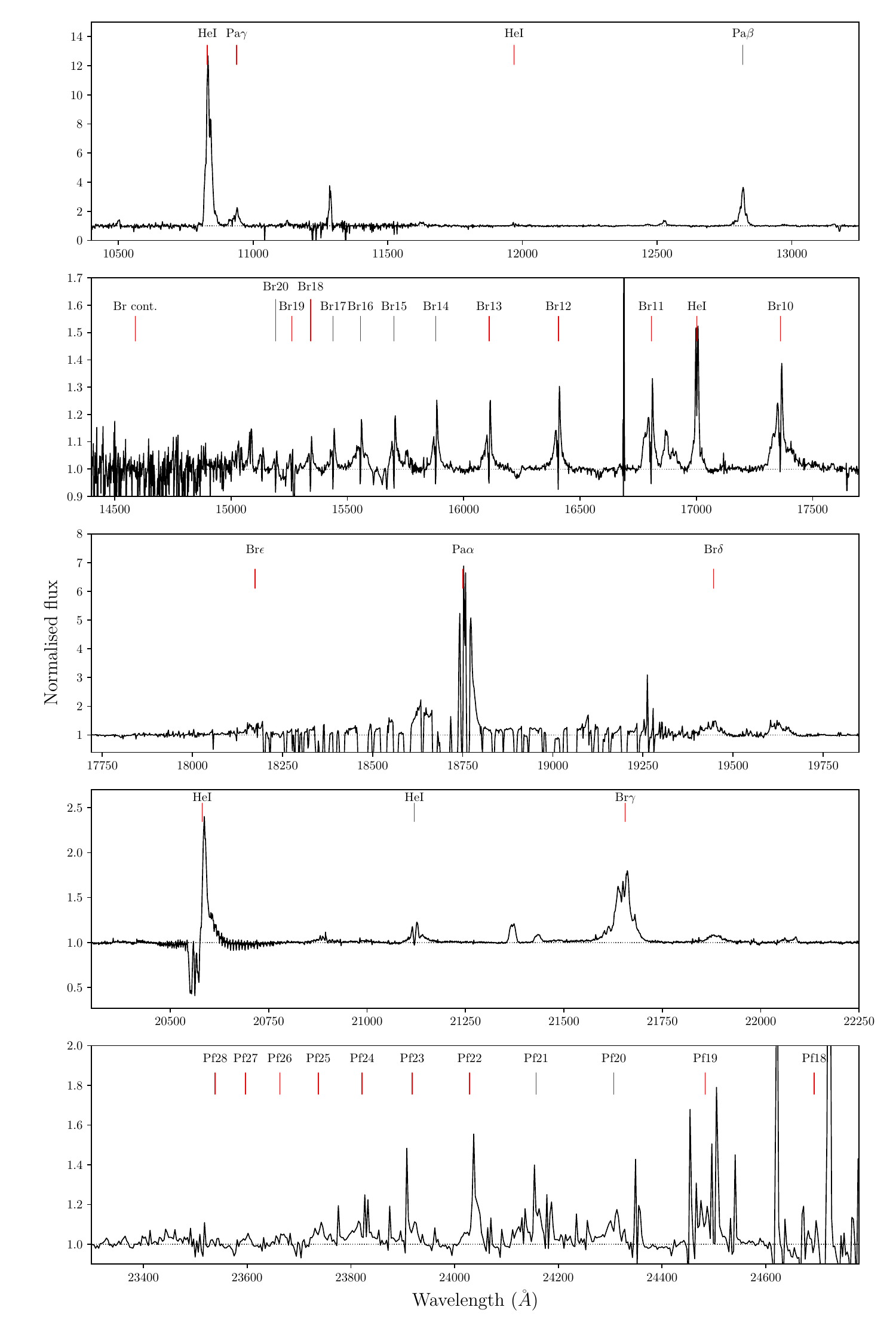}
    \caption{Full normalised spectrum of VLT/X-shooter epoch E-23D. \textit{Middle panel}: Pa$\alpha$ in the center alongside with Br$\delta$ and Br$\epsilon$. \textit{Bottom panel}: Higher orders of Pfund Series.}
    \label{figfullNIR23D}
\end{figure*}

\begin{figure}[ht]
    \centering
    \includegraphics[width=9truecm]{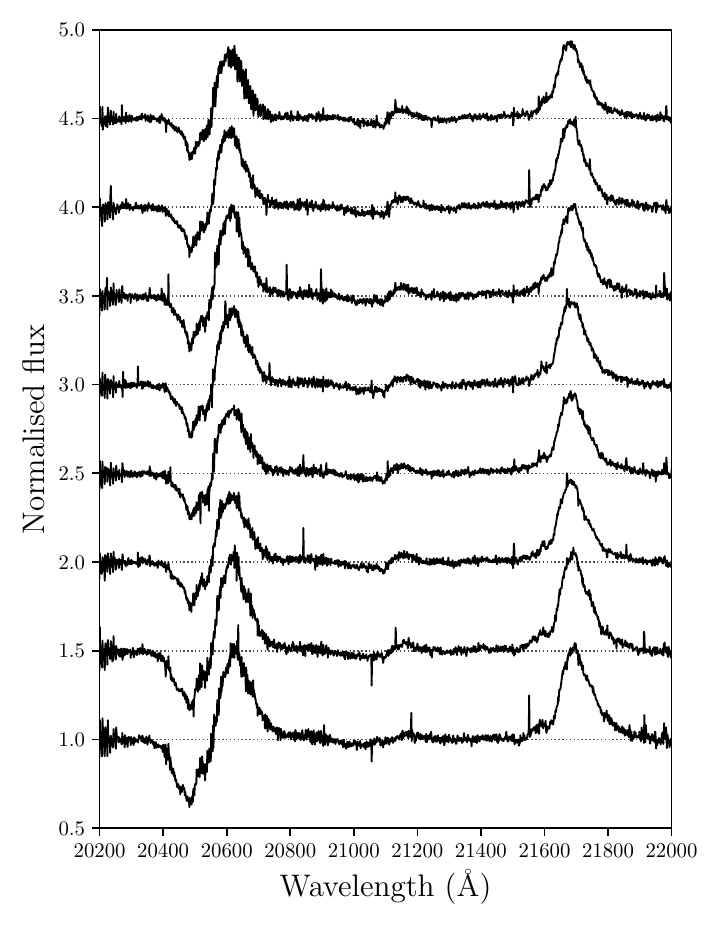}
    \caption{K-band region of epoch E-19C (VLT/X-shooter) with high time resolution (5 min each). The epoch has been divided into the eight spectra shown in the figure.}
    \label{figE19CHighTimeRes}
\end{figure}

\end{appendix}

\end{document}